\def\etal{{\it et al.\/\ }}

\def\sun{$_\odot$}

\def\arcsec{^{\prime\prime}}
\documentstyle[11pt,aaspp4,epsf]{article}
\lefthead{GARNETT ET AL.}
\righthead{CARBON IN I ZW 18}
\slugcomment{To appear in the Astrophysical Journal}
\begin{document}
\title{HIGH CARBON IN I ZWICKY 18: NEW RESULTS FROM 
{\it HUBBLE SPACE TELESCOPE} SPECTROSCOPY\altaffilmark{1}}
\author{D. R. Garnett\altaffilmark{2,} E. D. Skillman\altaffilmark{2}, 
R. J. Dufour\altaffilmark{3}, and G. A. Shields\altaffilmark{4}}
\altaffiltext{1}{Based on observations with the NASA/ESA Hubble Space
Telescope obtained at Space Telescope Science Institute, which is operated
by the Association of Universities for Research in Astronomy, under NASA
contract NAS5-26555.}
\altaffiltext{2}{Astronomy Department, University of Minnesota, 116 Church St., S. E., 
Minneapolis MN 55455; e-mail: garnett@oldstyle.spa.umn.edu, skillman@astro.spa.umn.edu}
\altaffiltext{3}{Department of Space Physics and Astronomy, Rice University, Houston, 
TX 77251-1892; e-mail: rjd@spacsun.rice.edu}
\altaffiltext{4}{Astronomy Department, University of Texas, Austin, TX 78712;
e-mail: shields@astro.as.utexas.edu}
\begin{abstract}
We present new measurments of the gas-phase C/O abundance ratio in both the 
NW and SE components of the extremely metal-poor dwarf irregular galaxy 
I~Zw~18, based on ultraviolet spectroscopy of the two H II regions using 
the Faint Object Spectrograph on the {\it Hubble Space Telescope}. We 
determine values of log C/O = $-$0.63$\pm$0.10 for the NW component and 
log C/O = $-$0.56$\pm$0.09 for the SE component. In comparison, log 
C/O = $-$0.37 in the sun, while log C/O = $-$0.85$\pm$0.07 in the three 
most metal-poor irregular galaxies measured by Garnett \etal (1995a). Our 
measurements show that C/O in I~Zw~18 is significantly higher than in 
other comparably metal-poor irregular galaxies, and above predictions
for the expected C/O from massive star nucleosynthesis. These results 
suggest that carbon in I~Zw~18 has been enhanced by an earlier population 
of lower-mass carbon producing stars; this idea is supported by stellar 
photometry of I~Zw~18 and its companion, which demonstrate that the current 
bursts of massive stars were not the first. Despite its very low metallicity, 
it is likely that I Zw 18 is not a ``primeval'' galaxy. 

\end{abstract}
\keywords{Galaxies: abundances -- galaxies: evolution -- galaxies: irregular --
galaxies: individual: I~Zw~18 -- galaxies: ISM -- H II regions} 

\section{Introduction}

The dwarf emission-line galaxy I~Zw~18 plays an important role in studies 
of the properties and evolution of dwarf irregular galaxies because of 
its extreme properties. Searle \& Sargent (1972) determined early on that 
I~Zw~18 (and its cousin II Zw 40) had a very low oxygen abundance; since 
then the O/H value has been refined to the current best estimate of 
(1.47$\pm$0.15)$\times$10$^{-5}$ in the NW H~II region (Skillman \& Kennicutt 
1993; hereafter SK93), the lowest value measured in any emission-line galaxy. 
The colors of I~Zw~18 are very blue, dominated by young massive stars, and 
no evidence has been found to date for an older red stellar population
(see Hunter \& Thronson 1995).

The oxygen deficiency and blue colors led to the conclusion that I~Zw~18
and similar galaxies are either very young, experiencing their first 
episodes of star formation, or have experienced only sporadic episodes
of star formation over their history (Searle, Sargent, \& Bagnuolo 1973). 
Meanwhile, surveys for other very metal-poor emission-line dwarf galaxies
(Kunth \& Joubert 1985, Terlevich et al. 1991) have failed to find any
more metal-poor than I Zw 18. This failure led Kunth \& Sargent (1986) 
to postulate that the H~II regions in I~Zw~18 were ``self-polluted'' by 
oxygen-rich ejecta from supernovae that had exploded during the lifetime
of the H~II region, and they proposed that the surrounding neutral gas
should be essentially pristine material (but see Tenorio-Tagle 1996 for
objections to this scheme).

If the ionized gas in I~Zw~18 is pristine material which has been contaminated
by Type II supernova ejecta, the heavy element abundance ratios should show
values characteristic of massive star nucleosynthesis. In particular, the gas
should be relatively deficient in elements produced mainly in low- and 
intermediate-mass stars, such as carbon and nitrogen, compared to oxygen,
which is produced in massive stars alone. Yet Dufour, Garnett, \& Shields
(1988) derived surprisingly high (essentially solar) values for C/O and N/O 
in I~Zw~18 from IUE UV and ground-based optical spectroscopy (since revised 
downward by Dufour \& Hester 1990 and SK93), suggesting that I~Zw~18 did 
experience contamination from an older generation of stars. On the other 
hand, Kunth et al. (1994) inferred from high-resolution GHRS spectra an 
oxygen abundance in the neutral gas that was much smaller than in the H II 
region in I~Zw~18 , and concluded that the ionized gas has been enriched by 
the present starburst (but see Pettini \& Lippman 1995 for a criticism of 
their analysis). Thus, there remains conflicting evidence over the nature
of I~Zw~18.

Because of the intimate connection between the production of He, C, and N 
in stars, accurate measurements of C, N, and O abundances in metal-poor 
galaxies are vital for determining the primordial He abundance; ultimately, 
one might expect that a model for the metallicity evolution of a galaxy 
combined with observed CNO abundances will provide the best estimates for 
the stellar He contamination and thus an accurate value for $Y_p$ (Balbes,
Boyd, \& Mathews 1993). In this 
paper, we report on new measurements of the C/O abundance ratio in 
I~Zw~18 obtained with the {\it Hubble Space Telescope (HST)}, and show that 
C/O in I~Zw~18 is significantly higher than observed in other very metal-poor 
dwarf galaxies, suggesting that I~Zw~18 has been enriched in carbon from an 
earlier generation of intermediate-mass stars.

\section{Observations and Analysis}

We observed both the NW and SE components of I~Zw~18 with the Faint Object 
Spectrograph on {\it HST}, using the red Digicon. The NW H II region was 
observed with the 0.86$\arcsec$ square (C-1) aperture and gratings G190H 
(exposure time 3600s) and G570H (300s) in Cycle 4, while the SE region was 
observed in Cycle 5 with the 0.86$\arcsec$ circular (B-3) aperture and 
gratings G190H (5170s), G570H (540s), and G400H (600s). Figure 1 (Plate 
XXX) shows the locations of the apertures superposed on H$\alpha$ and
F555W images of the galaxy taken with WFPC2. The spectra for the SE region 
were processed using the standard FOS pipeline reduction routines. The NW 
region spectra were reprocessed using the flat-fields and calibration 
observations taken through the C-1 apertures in July 1994; comparison of 
the pipeline and reprocessed spectra showed differences in relative 
photometry of no more than a few percent. The FOS spectra of the SE knot
are displayed in Figure 2.

We measured fluxes for the emission lines by direct integration of the 
emission line profiles. Interstellar reddening was estimated from 
H$\alpha$/H$\beta$ ratios measured from the G570H spectra; we measured values
for C(H$\beta$) = log F$_{H\beta}${\it (intrinsic)} $-$ log F$_{H\beta}${\it (observed)}
of 0.10$\pm$0.02 for the NW position and 0.03$\pm$0.04 for the SE position.
(These compare with the values 0.10$\pm$0.05 and 0.20$\pm$0.05 measured by SK93
for the NW and SE regions, respectively, suggesting that there is variable 
reddening within I Zw 18.)
The observed line fluxes were corrected for reddening using these values for
C(H$\beta$) with the average Galactic interstellar reddening curve of Seaton 
(1979), assuming all of the reddening is foreground. If one assumes the reddening
is internal, and that an SMC-like reddening law applies, the C~III]/H$\beta$
ratio in the NW region is reduced by approximately 10\%; the fluxes for the
SE region would be essentially unchanged. We list our reddening-corrected 
line fluxes for both regions in Table 1. 

The 1$\sigma$ uncertainties in the line fluxes in Table 1 were determined by 
adding in quadrature the error contributions from the following sources: the 
statistical noise in the lines plus local continuum, determined from the raw 
counts; the uncertainty in the photometric calibration of the FOS, approximately 
3\% (Bohlin 1995); and the error due to the uncertainty in the reddening correction.
Statistical noise dominates the uncertainty in the C~III] line flux.

To determine C and O abundances from the spectra, we adopted electron temperatures
and densities for the two regions determined by SK93, listed in Table 1. Garnett 
et al. (1995; G95a) were able to measure lines from both C~III] and O~III] 
$\lambda$1666 in other H~II regions, and thus to derive C$^{+2}$/O$^{+2}$ abundance 
ratios with only small uncertainties from reddening and temperature errors. Here 
we were not able to measure O~III] $\lambda$1666, however, so we must compare the 
C~III] emission with the optical [O~III] lines to determine the C$^{+2}$/O$^{+2}$ 
ratio.\footnotemark[5] 
\footnotetext[5]{Our 2$\sigma$ upper limit for O~III] $\lambda$1666 leads to
a lower limit of log C/O $\approx$ $-$1.0, not a very strong constraint.}
We computed level populations and line emissivities $\epsilon$($\lambda$) using 
a five-level atom program and our adopted electron temperatures, assuming the 
same temperature applies to the C$^{+2}$ and O$^{+2}$ emitting regions (Garnett 
1992). Ionic abundances relative to hydrogen then follow from  
\begin{equation}
{{C^{+2}\over O^{+2}}} = 
{\epsilon(\lambda5007)\over \epsilon(\lambda1908)}*{I(\lambda1908)\over I(\lambda5007)}. 
\end{equation}

As in G95a, we apply a small ionization correction (ICF) to convert the measured 
C$^{+2}$/O$^{+2}$ ratios to final C/O abundance ratios. We estimated these 
corrections as in G95a (see that paper, especially Figure 2,
for details), based on the measured O$^+$/O ratios and the ionization fractions 
computed from a photoionization model grid; these corrections were less than 
10\% for both regions. Our final values for C/O in the two components of I~Zw~18 
are listed in Table 1. The uncertainties in the final abundances were estimated 
by summing in quadrature the contributions from the line flux measurements, the 
uncertainty in the ratio of the C III] and [O III] emissivities from errors in 
T$_e$, and the uncertainty in the ICFs. Moderate density fluctuations have a 
negligible effect on C$^{+2}$/O$^{+2}$ derived this way; based on a five-level 
atom calculation, an average electron density of order 10$^6$ cm$^{-3}$ within 
the 0.$\arcsec$86 FOS aperture is required to reduce the C/O ratio by 0.2-0.3 
dex through collisional de-excitation of the optical [O~III] lines. There is no 
evidence for such dense gas from the H$\alpha$ emission measure or spectroscopic
diagnostics.

\section{Discussion}

Figure 3 shows our new results for C/O in I~Zw~18 compared with the data from 
G95a plus new observations of NGC 5253 by Kobulnicky et al. (1997). G95a noted
that their data (unfilled circles in Fig. 2, plus a lower limit for I~Zw~18,
not shown) showed a monotonic increase in C/O with increasing O/H; note that
the Kobulnicky et al. (1997) measurements of C/O in three positions in NGC 5253 
(unfilled triangles) fall along the same trend as the G95a points. A fit to the
nine dwarf galaxy measurements (excluding I~Zw~18) give a slope of 0.41$\pm$0.07
in log C/O vs log O/H. The dispersion about this fit is surprisingly small: the
rms dispersion from the power-law fit is only $\pm$0.05 dex. However, we can not
say yet what the true relation is from such a small sample; for example, in halo 
stars C/O levels off for log O/H $<$ $-$4 (see G95a, Fig. 5). Second, any 
intrinsic dispersion should be largest at the lowest metallicities, where a single 
starburst can have a greater effect. At higher O/H, the intrinsic dispersion could 
be quite small. Small number statistics inhibit any interpretation of the measured
dispersion in C/O. Until enough C/O measurements are obtained to define the trend
reliably, we shall rely on comparison with the most metal-poor galaxies to interpret
the I~Zw~18 measurements.

The three most metal-poor galaxies from the G95a sample have a mean log C/O = 
$-$0.85$\pm$0.07 (mean error); by comparison, the mean log C/O = $-$0.60$\pm$0.09 
in I~Zw~18 is 2.9$\sigma$ above this average (and even higher than expected from 
extrapolating the trend of C/O vs O/H observed by G95a).  Similarly, the mean 
log C/N = +0.98 in I~Zw~18 is also well above the values determined for the other 
very metal-poor dwarf galaxies (G95a).

Are there systematic effects that could artificially enhance C/O in I~Zw~18?
We examine several possibilities:

(1) {\it Errors in electron temperature}. A systematic error in T$_e$ can affect 
the derived C$^{+2}$/O$^{+2}$, because of the large difference in excitation
(C$^{+2}$/O$^{+2}$ varies as $e^{4.65/t}$, where t = T$_e$/10$^4$ K). However,
to reduce the observed C/O in I~Zw~18 to log C/O = $-$0.85 would require that 
T$_e$ be much higher than measured: approximately 25,000 K for the NW region and
23,000 K for the SE region. These are very high temperatures for photoionized
nebulae, difficult to produce even in H II regions with extremely low abundances. 
Note that in the case of temperature fluctuations (Peimbert 
1967), the average electron temperature is expected to be smaller than observed,
so we would derive a higher C/O than given. One possibility is that the C~III]
lines are formed in a region of significantly higher T$_e$ than the optical 
[O~III] lines, although photoionization models do not predict such large
differences (Garnett 1992).

(2) {\it Depletion onto grains}. G95a discussed depletion of C and O onto grains 
at length. It is expected that C would be more depleted than O in the ISM (see 
Mathis 1996), but the degree to which either element is depleted in
the ISM is highly uncertain. One could argue that the galaxies with low C/O 
have more C in grains, but one would need to ask why carbon depletions vary 
by a factor of two in otherwise similar starbursts. Furthermore, these galaxies
have very similar Si/O abundance ratios (Garnett et al. 1995b), raising the 
question of why C would suffer large depletion variations while Si does not.

(3) {\it Interstellar reddening}. C/O could be reduced if the interstellar 
reddening is overestimated. However, the measured interstellar reddening in 
I~Zw~18 is so low that this can contribute no more than 10-15\% to the high 
C/O values. Errors in reddening are not a problem for the G95a observations.

(4) {\it Ionization}. Ionization effects can reduce C/O by no more than 10-20\%, 
not enough to account for the higher values in I~Zw~18. To raise C/O in the very
low abundance G95a galaxies would require that most of the carbon be in the
form of either C II or C IV, neither of which is observed to be strong in
such galaxies (from IUE observations), and such extreme ionization conditions
would be at odds with the ionization inferred from the optical data.

This discussion shows that the most likely systematic effects tend to lead
to higher C/O values than we have derived for I~Zw~18. A deeper UV spectrum 
to detect the O III] $\lambda$1666 line would yield a more definitive measurement 
of C/O in I~Zw~18. However, the analysis above and the agreement between the
two independent measurements lead us to conclude that C/O 
in I~Zw~18 is elevated compared to other very oxygen-poor irregular galaxies. 

An elevated C/O abundance ratio in I~Zw~18 would suggest that the galaxy
has experienced contamination by carbon from an older generation of stars,
and that the current burst of star formation is not the first. G95a noted 
that nucleosynthesis models for 10-42 M$_{\sun}$ stars by Weaver \& Woosley 
(1993) predict an integrated log C/O = $-$0.83 for their ``best estimate'' of 
the $^{12}$C($\alpha,\gamma$)$^{16}$O nuclear reaction rate, corresponding 
quite well with the values measured by G95a in their most oxygen-poor galaxies.
If this correctly represents the nucleosynthesis contribution from massive
stars alone, then the additional carbon observed in two widely separated 
locations in I~Zw~18 must come from lower mass, long-lived carbon star
and planetary nebula progenitors with lifetimes $>>$ 10 Myr.

A model by which this situation may come about at extremely low O/H is one
in which the evolution of a dwarf galaxy is influenced by the effects of 
discrete starbursts (Matteucci \& Tosi 1985, Garnett 1990, Pilyugin 1993). 
In such models, heavy elements that are produced in massive stars, such as 
oxygen, are injected into the ISM by a starburst early and can enrich the 
gas on relatively short timescales; supernova-driven galactic winds may also 
eject much of the oxygen-rich SN ejecta into the galaxy halo and reduce the 
effective oxygen yield. After the massive stars have died away, lower-mass 
stars can enrich the ISM in carbon (and primary nitrogen) in more gentle 
mass loss events, leading to higher C/O and N/O than one would expect in a 
young galaxy which has been enriched by massive stars alone. The relative 
abundances of C, N, and O then provide an indication of the time elapsed 
since the last major starburst in a dwarf galaxy (Edmunds \& Pagel 1978).

In Figure 4 we compare the observed C/O ratio in I~Zw~18 with the predictions
of chemical evolution models for blue compact dwarf galaxies as a function 
of galaxy age. The filled squares show models computed by Kunth, Matteucci,
\& Marconi (1995; KMM) specifically to explain the abundances in I~Zw~18. 
These are all one-burst models of short age, except for their model 6, which 
is a model consisting of two bursts, one that occured 1 Gyr ago and one 
beginning only 10 Myr ago. We show their predicted abundances for I~Zw~18 
at the beginning of the second burst, after 990 Myr. We also show the 
results of general models for metal-poor blue compact dwarf galaxies by 
Carigi et al. (1995; CCPS). These models were computed assuming continuous 
star formation rather than starbursts, and different massive star yields. 
Both sets of models include the effects of differential (heavy element 
enriched) winds. The hatched region shows the values of C/O encompassed 
by the observations of I~Zw~18 and corresponding errorbars.

None of the KMM models are able to account for the high C/O in I~Zw~18,
suggesting that an additional source of carbon is needed in their models.
The CCPS models can explain our observed C/O ratios for galaxy ages of
the order 1 Gyr. From these results it could be inferred that I~Zw~18 
had an episode of star formation that occured a few hundred million years 
ago that led to the presently observed levels of carbon and nitrogen. 
Interestingly, this corresponds roughly to the age inferred from stellar
photometry in the companion irregular galaxy NW of the main body of 
I~Zw~18 (Dufour et al. 1996), which may suggest that an older population
may still lay obscured by the light of the luminous present-day burst
in the main body.

The agreement in N/O and O/H between the NW and SE components of I~Zw~18
led SK93 to question the validity of the ``self-pollution'' model for
I~Zw~18. The agreement between our two high values for C/O further 
strengthen this argument. It is highly improbable that these two separate 
star formation events could lead to essentially identical abundances in a
self-pollution model.

We caution that C/O and N/O abundance measurements alone are insufficient 
to constrain the star formation history of a dwarf galaxy, although they
can set some constraints on the bursting model. There is still significant 
uncertainty in the yields of C and N due to uncertainties in the 
$^{12}$C($\alpha,\gamma$)$^{16}$O rate (cf. Buchmann 1996, Trautvetter 
et al. 1996, and France et al. 1996), and in the effects of convective 
mixing. A larger sample of high-quality C/O measurements for low abundance 
galaxies should tell us if there is significant scatter in C/O at fixed 
O/H, and thus a significant spread in the time since previous starbursts. 
However, evidence is growing that the most metal-poor dwarf galaxies may 
not be primordial after all.

\acknowledgments
We thank Jen Christensen of STScI for assistance in reprocessing the FOS 
spectra, Chip Kobulnicky for providing results prior to publication, and 
the referee, Elena Terlevich, for suggestions which helped to 
improve the text. Support for this program was provided by NASA through 
grants GO-5434-93A and GO-6536-94A from STScI. EDS acknowledges support 
from NASA-LTSARP grant NAGW-3189 and a Bush Sabbatical Fellowship from the 
graduate school of the University of Minnesota and would like to thank the 
Max-Planck-Institute for Astrophysics for hospitality during his sabbatical 
year. GAS gratefully acknowledges a Faculty Research Assignment from the 
University of Texas at Austin, and the hospitality and support of Rice 
University, during spring 1995.

\newpage

\begin{deluxetable}{lccc}
\tablewidth{35pc}
\tablecaption{I Zw 18 FOS Spectra and Derived Properties}

\tablehead{
\colhead {Quantity}  & \colhead{NW Region} &\colhead{SE Region}& \colhead{} }
\startdata
I(O III] 1666)     & $<$ 0.27 (2$\sigma$)   & $<$ 0.29 (2$\sigma$) &   \nl
I(C III] 1909)     & 0.55 $\pm$ 0.09        & 0.47 $\pm$ 0.07      &   \nl
I([O II] 3727)     &   \nodata              & 0.32 $\pm$ 0.04      &   \nl
I(H$\delta$) 4101) &   \nodata              & 0.24 $\pm$ 0.03      &   \nl
I(H$\gamma$) 4340) &   \nodata              & 0.50 $\pm$ 0.02      &   \nl
I(H$\beta$) 4861)  & 1.00 $\pm$ 0.04        & 1.00 $\pm$ 0.03      &   \nl
I([O III] 4959)    & 0.78 $\pm$ 0.04        & 0.70 $\pm$ 0.04      &   \nl
I([O III] 5007)    & 2.02 $\pm$ 0.08        & 2.18 $\pm$ 0.04      &   \nl
I(H$\alpha$ 6563)  & 2.76 $\pm$ 0.10        & 2.77 $\pm$ 0.05      &   \nl
                   &                        &                      &   \nl
C(H$\beta$)        & ~0.10 $\pm$ 0.02       & ~0.03 $\pm$ 0.03     &   \nl
n$_e$              & 100                    & 100                  &   \nl
T(O III) (K)\tablenotemark{a}   & 19,600 $\pm$ ~900      & 17,200 $\pm$ 1200  &   \nl
T(O II) (K)\tablenotemark{a}    & 15,300 $\pm$ 1000      & 14,500 $\pm$ 1300  &   \nl
\tablenotetext{a}{Adopted from Skillman \& Kennicutt 1993}
                   &                        &                      &   \nl
Log O/H            & $-$4.83 $\pm$ 0.04\tablenotemark{a} & $-$4.74 $\pm$ 0.05\tablenotemark{a} & \nl
O$^+$/O            & ~~0.16 $\pm$ 0.03\tablenotemark{a}  & ~~0.13 $\pm$ 0.03  &  \nl
C$^{+2}$/O$^{+2}$  & ~~0.22 $\pm$ 0.05       & ~~0.25 $\pm$ 0.05       &  \nl
ICF                & ~~1.06 $\pm$ 0.11       & ~~~1.09 $\pm$ 0.14      &  \nl
Log C/O            & $-$0.63 $\pm$ 0.10      & $-$0.56 $\pm$ 0.09      &  \nl
Log C/N            & +0.93 $\pm$ 0.13        & +1.04 $\pm$ 0.11        &  \nl
\enddata
\end{deluxetable}

\newpage

\begin{figure}
\plotfiddle{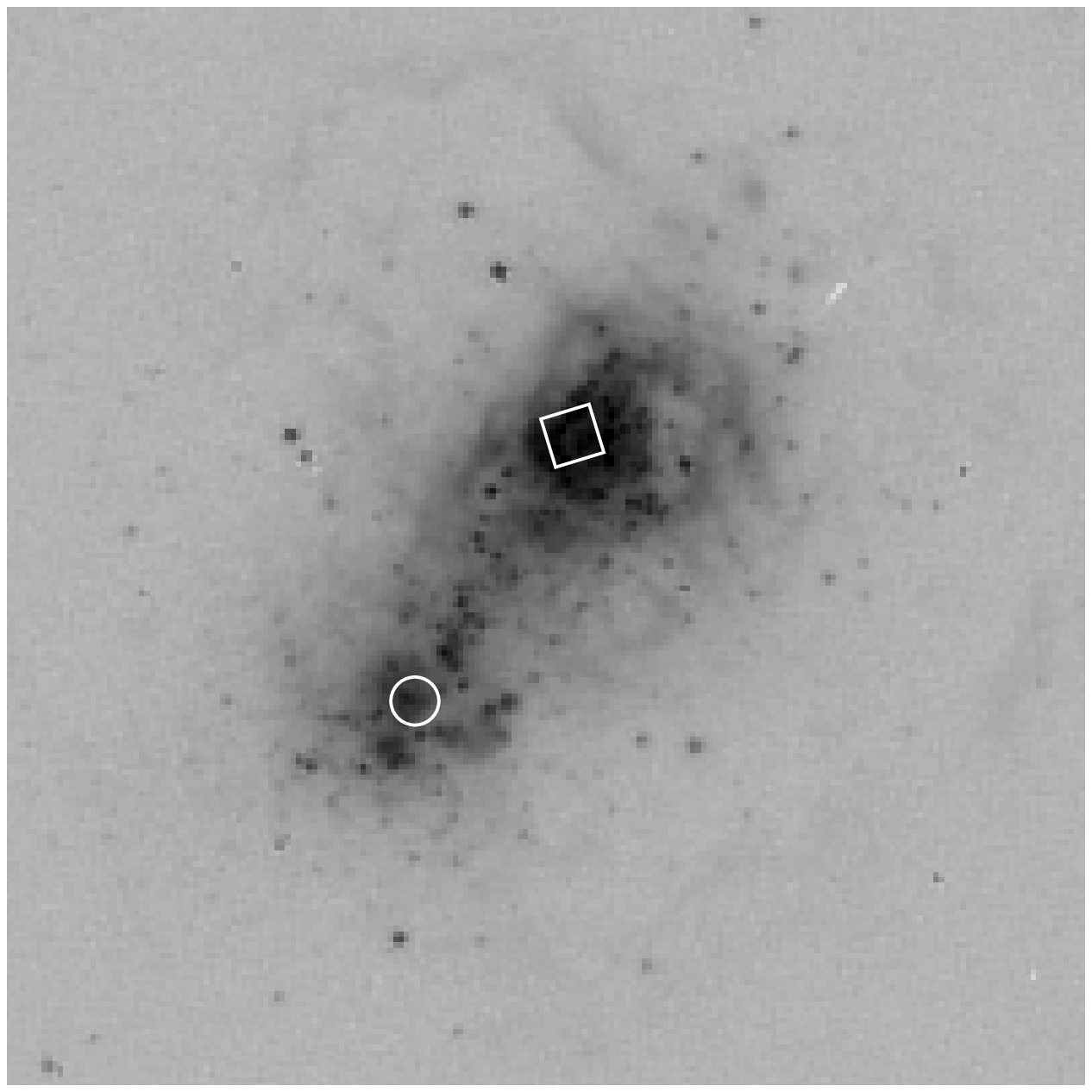}{3.5in}{00}{70}{70}{-210}{-140}
\plotfiddle{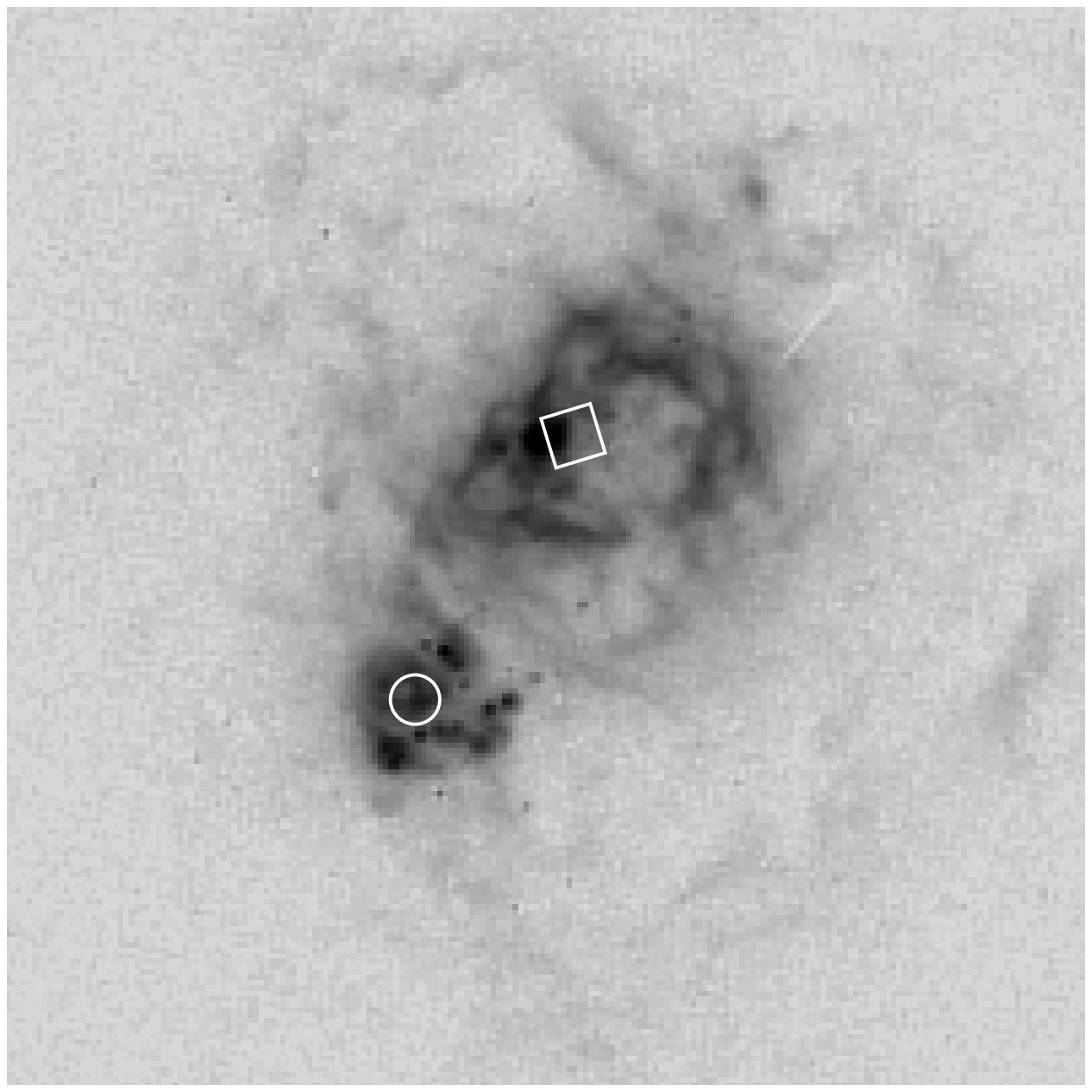}{3.5in}{00}{70}{70}{-215}{-140}
\caption{Portions of WFPC2 F656N (bottom; total exposure 4600s) and F555W 
(top; exposure 4600s) images of I~Zw~18, showing the locations of the 
apertures for our FOS spectra. Coordinates for the aperture locations 
were (J2000): 
9$^h$ 34$^m$ 2$^s$.32, +55$^{\deg}$ 14$^{\prime}$ 22$^{\prime\prime}$.7 
for the SE region, and 
9$^h$ 34$^m$ 2$^s$.04, +55$^{\deg}$ 14$^{\prime}$ 28$^{\prime\prime}$.0 
for the NW region. North is up and east is to the left.  }
\end{figure}

\newpage

\begin{figure}
\plotfiddle{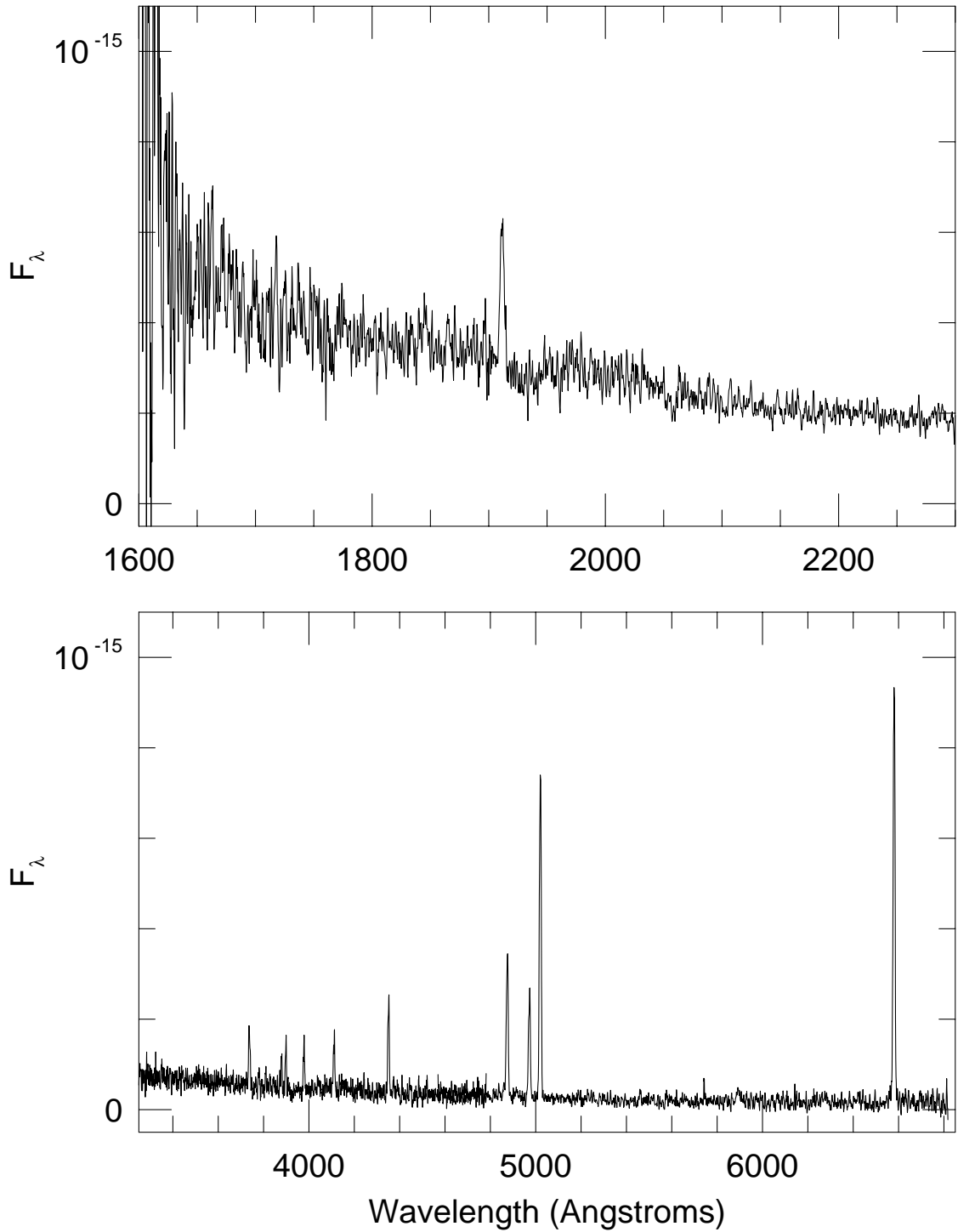}{8in}{0}{80}{80}{-245}{00}
\caption{FOS spectrum of the SE component of I~Zw~18. The top
panel shows the G190H spectrum, bottom panel the combined G400H and G570H
spectra. The spectra have been smoothed with a three-point boxcar. Note
the clear detection of C~III] $\lambda$1908.}
\end{figure}

\newpage

\begin{figure}
\plotfiddle{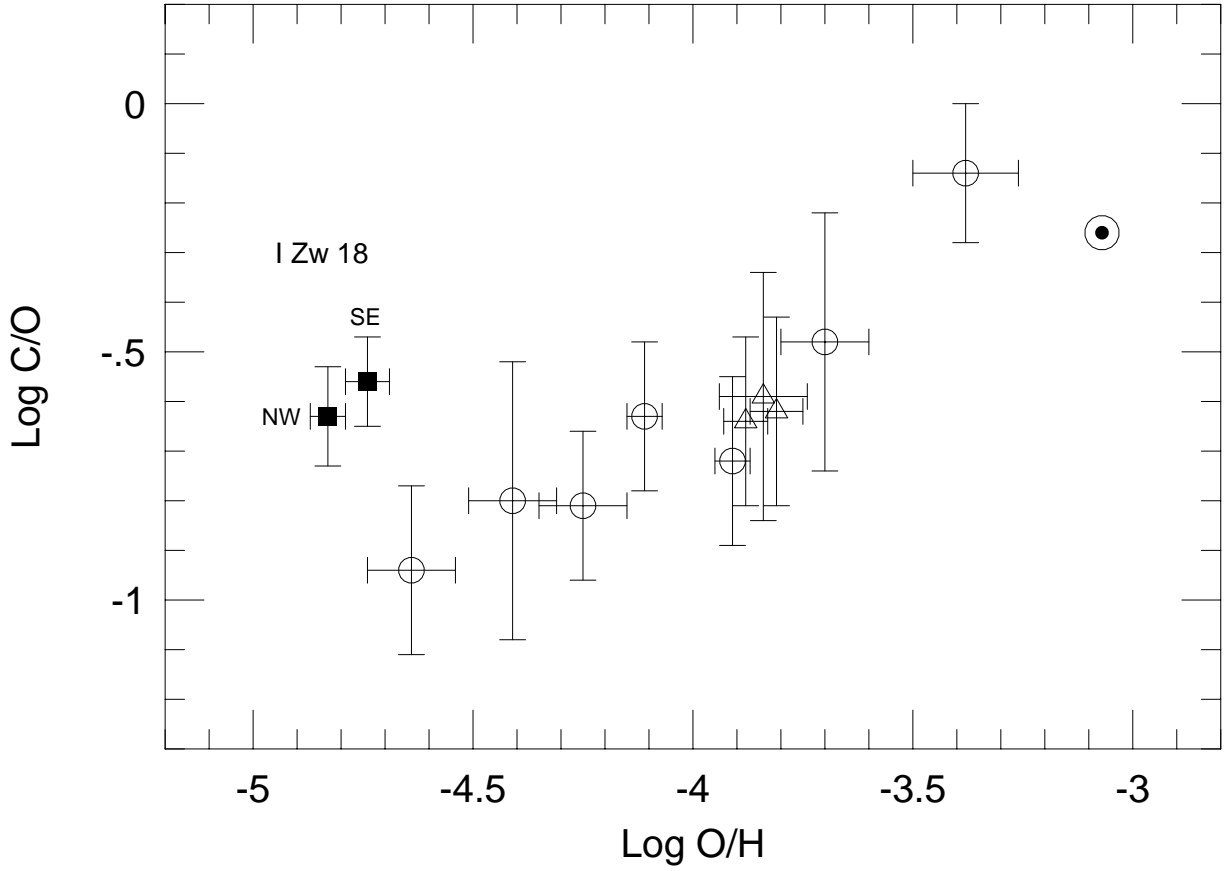}{6in}{90}{65}{65}{260}{100}
\caption{Log C/O vs Log O/H from FOS spectroscopy of I~Zw~18 (filled
squares), compared with data from Garnett et al (1995a; unfilled circles)
and Kobulnicky et al. (1996; unfilled triangles). Solar value is from
Grevesse \& Noels (1993).}
\end{figure}

\newpage

\begin{figure}
\plotfiddle{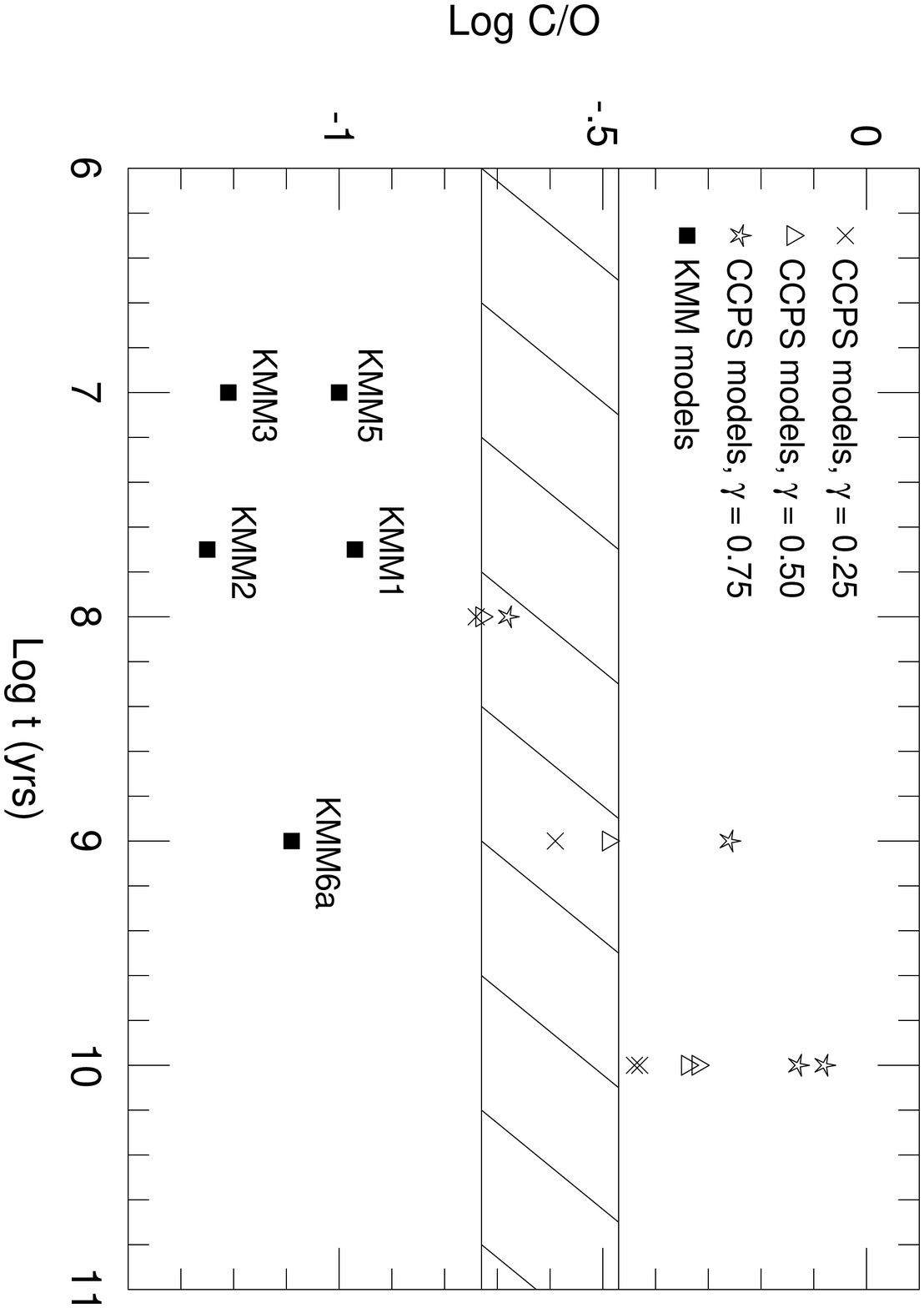}{6in}{90}{65}{65}{260}{100}
\caption{Log C/O vs. log age from chemical evolution models for 
I~Zw~18 (KMM) and blue compact dwarf galaxies (CCPS). The KMM models are 
labeled according to model number; KMM6a shows abundances from the KMM 
two-burst model 6 after 990 Myrs, at the beginning of the second burst. For 
the CCPS models, $\gamma$ refers to the strength of the SN-driven galactic 
wind. The hatched area shows the range occupied by the observed values 
of C/O in I~Zw~18 and corresponding errors.}
\end{figure}


\clearpage

\begin{references}

\reference{b1}
Balbes, M. J, Boyd, R. N., \& Mathews, G. J., 1993, ApJ, 418, 229

\reference{b2}
Bohlin, R. C., 1995, in Calibrating Hubble Space Telescope: Post Servicing Mission
eds. A. Koratkar and C. Leitherer (Baltimore: STScI), p. 49

\reference{b3}
Buchmann, L., 1996, Nucl. Phys. A, in press

\reference{c1}
Carigi, L., Col\' \i n, P., Peimbert, M., \& Sarmiento, A., 1995, ApJ, 445, 98 (CCPS)

\reference{d1}
Dufour, R. J., Garnett, D. R., \& Shields, G. A. 1988, ApJ, 332, 752

\reference{d2}
Dufour, R. J., Garnett, D. R., Skillman, E.D., \& Shields, G. A. 1996, in
From Stars to Galaxies: The Impact of Stellar Physics on Galaxy Evolution
eds. C. Leitherer, U. Fritze-von Alvensleben, and J. Huchra, ASP Conf. Series
98, p. 358

\reference{d3}
Dufour, R. J., \& Hester, J. J., 1990, ApJ, 350, 149

\reference{e1}
Edmunds, M. G., \& Pagel, B. E. J., 1978, MNRAS, 185, 78P

\reference{f1}
France, R. H. III, Wilds, E. L., Jevtic, N. B., McDonald, J. E., \& Gai, M. 
1996, Nucl. Phys. A, in press

\reference{g1}
Garnett, D. R., 1990, ApJ, 363, 142 

\reference{g2}
Garnett, D. R., 1992, AJ, 103, 1330 

\reference{g3}
Garnett, D. R., Skillman, E. D., Dufour, R. J., Peimbert, M., Torres-Peimbert, S.,
Terlevich, R., Terlevich, E., \& Shields, G. A., 1995a, ApJ, 443, 64 (G95a)

\reference{g4}
Garnett, D. R., Dufour, R. J., Peimbert, M., Torres-Peimbert, S., Shields, G. A., 
Skillman, E. D., Terlevich, E., \& Terlevich, R. J., 1995b, ApJ, 449, L77 

\reference{g5}
Grevesse, N., and Noels, A. 1993, in Origin and Evolution of the Elements,
eds. N. Prantzos, E. Vangioni-Flam, \& M. Casse (Cambridge: Cambridge
University Press), 15

\reference{h1}
Hunter, D. A., \& Thronson, H. A., 1995, ApJ, 452, 238

\reference{k1}
Kobulnicky, H. A., Skillman, E. D., Roy, J.-R., Walsh, J. R., \& Rosa, M. R., 1997, 
ApJ, in press 

\reference{k2}
Kunth, D., \& Joubert, M., 1985, A\&A, 142, 411 

\reference{k3}
Kunth, D., Lequeux, J., Sargent, W. L. W., \& Viallefond, F., 1994, A\&A, 282, 709 

\reference{k4}
Kunth, D., Matteucci, F., \& Marconi, G., 1995, A\&A, 297, 634 (KMM) 

\reference{k5}
Kunth, D., \& Sargent, W. L. W., 1986, ApJ, 300, 496 

\reference{m1}
Mathis, J. S. 1996, ApJ, in press

\reference{m2}
Matteucci, F., \& Tosi, M., 1985, MNRAS, 217, 391

\reference{p1}
Peimbert, M. 1967, ApJ, 150, 825

\reference{p2}
Pettini, M., \& Lipman, K. 1995, A\&A, 297, L63

\reference{p3}
Pilyugin, L. S., 1993, A\&A, 277, 42 

\reference{s1}
Searle, L. C., \& Sargent, W. L. W. 1972, ApJ, 173, 25

\reference{s2}
Searle, L. C., Sargent, W. L. W., \& Bagnuolo, W. G., 1973, ApJ, 179, 427

\reference{s3}
Seaton, M. J., 1979, MNRAS, 185, 57P

\reference{s4}
Skillman, E. D., \& Kennicutt, R. C., Jr., 1993, ApJ, 411, 655 (SK93)

\reference{t1}
Tenorio-Tagle, G. 1996, AJ, 111, 1641 

\reference{t2}
Terlevich, R. J., Melnick, J., Masegosa, J., Moles, M., \& Copetti, M. V. F., 
1991, A\&AS, 91, 285

\reference{t3}
Trautvetter, H. P., Roters, G., Rolfs, C., Schmidt, S., \& Descouvemont, P., 
1996, Nucl. Phys. A, in press

\reference{w1}
Weaver, T. A., \& Woosley, S. E. 1993, Phys. Rep., 227, 65
 
\end{references}
\end{document}